\documentclass[aps,prl,longbibliography,lengthcheck,superscriptaddress,nopreprintnumbers,reprint,groupedaddress]{revtex4-1}
\usepackage{natbib,units}
\usepackage{graphicx}
\usepackage{dcolumn}
\usepackage{bm}        
\usepackage{amssymb}   

\usepackage{bm}
\usepackage[usenames ,dvipsnames]{xcolor}
\usepackage{color}
\usepackage{url}
\usepackage{dsfont}
\usepackage{slashed}
\usepackage{epsfig}
\usepackage{multirow,array}
\usepackage{float}
\usepackage{fancyhdr}
\usepackage{indentfirst}
\usepackage{cancel}
\usepackage{tabularx}
\usepackage{simplewick}
\usepackage{amsmath}
\usepackage[normalem]{ulem}
\usepackage{slashed}
\usepackage{upgreek}
\usepackage{xtab,afterpage,longtable}
\RequirePackage[colorlinks=true
  ,urlcolor=Green
  ,anchorcolor=blue
  ,citecolor=orange
  ,filecolor=blue
  ,linkcolor=blue
  ,menucolor=blue
  ,pagecolor=blue
  ,linktocpage=true
  ,pdfproducer=medialab
  ,pdfa=true
]{hyperref}

\newcommand{\mnras}{MNRAS}

\newcommand{\jcap}{JCAP}
\newcommand{\physrep}{\textit{Phys.~Rep.}}

\begin{document}

\title{Diversifying halo structures in two-component self-interacting dark matter models via mass segregation}

\author{Daneng Yang}
\email{yangdn@pmo.ac.cn}
\author{Yue-Lin Sming Tsai}
\email{smingtsai@pmo.ac.cn}
\author{Yi-Zhong Fan}
\email{yzfan@pmo.ac.cn}

\affiliation{Purple Mountain Observatory, Chinese Academy of Sciences, Nanjing 210023, China}
\affiliation{School of Astronomy and Space Sciences, University of Science and Technology of China, Hefei 230026, China}

\date{\today}

\begin{abstract}
Self-interacting dark matter (SIDM), through gravothermal evolution driven by elastic self-scatterings, offers a compelling explanation for the observed diversity of inner halo densities. In this work, we investigate SIDM dynamics in a two-component dark matter model with mass ratios of order unity, motivated by an asymmetric dark matter framework that naturally evades constraints from relic abundance and mediator decay, while enabling strong, velocity-dependent self-interactions. We show that cross-component scatterings significantly enhance mass segregation, driving the formation of dense, core collapsed-like halos. This effect couples naturally to SIDM-induced diversity, introducing a new mechanism for generating structural variations beyond those arising from gravothermal evolution alone. Our results reveal a novel mechanism for reconciling SIDM with small-scale observational tensions by enabling shifts in central densities while preserving the flexibility to generate diverse halo structures. We further highlight that halo structural diversity may serve as a diagnostic of dark sector composition, opening a new observational window into the particle nature of SIDM.
\end{abstract}

\maketitle

{\it Introduction.}
While the cold dark matter (CDM) paradigm has been remarkably successful in explaining the large-scale structure of the universe through gravitational interactions, the microscopic nature of dark matter, from its composition to interaction properties, remains fundamentally unknown. 
Uncovering its particle nature requires physics beyond the Standard Model and provides a unique window into early-universe processes, including the origin of the universe, baryogenesis, and the interplay between inflation and the emergence of cosmic structure~\cite{Cirelli:2024ssz,Arbey:2021gdg,Balazs:2024uyj}.
Among the various theoretical frameworks, self-interacting dark matter (SIDM) has emerged as a compelling candidate, particularly in addressing persistent discrepancies on galactic and sub-galactic scales~\cite{Bullock:2017xww,Salucci:2018hqu,Gentile:2004tb,vanDokkum:2022zdd,Bonaca:2018fek}. 
Elastic scattering among dark matter particles in SIDM models drives gravothermal evolution, leading to either cored or cuspy halo density profiles, depending on the interaction strength and the system's evolutionary stage~\cite{Tulin170502358,Adhikari220710638,Spergel9909386,kochanek:2000pi,kamada:2016euw,ren180805695,santos-santos191109116,correa:2022dey,yang:2022mxl,zavala:2019sjk,kaplinghat:2019svz,turner:2020vlf,slone:2021nqd,silverman:2022bhs,yang220503392,yang:2023jwn,Yang:2024tba,Hou:2025gmv,Robertson:2018anx,Chu:2018fzy,Chu:2018faw,Fischer:2020uxh,Fischer:2022rko,Zeng:2021ldo,Nadler:2025jwh}.
This mechanism provides a natural explanation for a variety of extreme and otherwise puzzling astrophysical observations, including dark matter-deficient galaxies, supermassive black holes at high redshifts, strong lensing perturbers, excessive small-scale lenses, and even the final-parsec problem in binary black hole mergers~\cite{Nadler:2023nrd,Kong220405981,Zhang:2024ggu,Zhang:2024qem,Zhang:2024qmh,sameie:2019zfo,Yang:2020iya,Kong:2025irr,balberg:2001qg,pollack:2014rja,choquette:2018lvq,feng:2021rst,meshveliani:2022rih,Alonso-Alvarez:2024gdz,Jiang:2025jtr}.

Although the non-luminous nature of dark matter renders its internal composition unobservable in principle, this degeneracy can be lifted in multi-component scenarios with cross-component interactions, as we demonstrate in this work.
The possibility of multiple dark matter species is not only a natural extension but is also well-motivated theoretically.
In minimal SIDM models involving a single species coupled to a light mediator, efficient annihilation channels often produce signals in conflict with Cosmic Microwave Background and indirect detection constraints~\cite{Bringmann:2016din}. These tensions motivate consideration of richer dark sector structures, such as asymmetric dark matter models~\cite{Kaplan:2009ag,Petraki:2011mv}. These models postulate two dark matter species with an asymmetry analogous to the baryon asymmetry in the visible sector~\cite{Kaplan:2009ag,Petraki:2011mv,Cyr-Racine:2012tfp,Petraki:2013wwa,Petraki:2014uza,ZUREK201491,Han:2023olf}. In the early universe, the symmetric components annihilate efficiently, leaving behind the asymmetric population that determines the relic abundance. Because the relic density is no longer determined by the annihilation cross section, these models naturally accommodate strong self-interactions.
Furthermore, just as the proton and electron differ in mass while interacting via the electromagnetic force, dark matter components can have distinct masses while coupling through a dark-sector interaction~\cite{Petraki:2014uza,Cyr-Racine:2012tfp}. This setup introduces novel structure formation dynamics, which we explore via cosmological $N$-body simulations of two-component SIDM. In particular, we focus on a mass ratio of $3:1$ between the species and demonstrate that inter-species interactions lead to significant mass segregation, resulting in a broader range of inner density profiles compared to single-component SIDM.

Prior simulations of two-component dark matter have typically focused on either large mass ratios, as in atomic dark matter, or near-degenerate states~\cite{Low:2025llz,Roy:2023zar,Schutz:2014nka}, without considering the effects of gravothermal evolution. 
Ref.~\cite{choquette:2018lvq} performed N-body simulations of a dark matter subcomponent that self-interacts, focusing on its gravothermal collapse as a potential seed for a supermassive black hole.
In contrast, our work focuses on the density profiles of entire dark matter halos, demonstrating how two-component SIDM reshapes their structural diversity.
Despite its simplicity, our framework is broadly applicable and offers a versatile mechanism for generating either diffuse or dense halo structures.
These features may help resolve outstanding tensions in SIDM interpretations of recent precision data, including observations of small-scale lenses, bullet clusters, halo ellipticities, and the Tully-Fisher relation~\cite{Correa:2024vgl,Kong:2024zyw,yang:2023stn,Ebisu:2021bjh,Vargya:2021qza,Correa:2020qam}.

{\it Cosmological simulation of a two component SIDM model.}
We simulate a dark QED-like model consisting of two dark matter components, $\chi_H$ and $\chi_L$, with a mass ratio $m_H/m_L = 3$ and equal number densities.
For comparison, we also simulate a model with a single dark matter component, $\chi_0$, using the same total number of particles and assigning it a mass $m_0 = 2m_H/3$ to match the total mass of the two-component system.
The scattering cross sections are modeled using the Møller and Rutherford equations parametrized by $\sigma_0/m$ and $w$, following Ref.~\cite{yang220503392}. We assume a fixed mediator mass and a dark fine-structure constant in the perturbative regime, which results in different $\sigma_0/m$ and $w$ values for each scattering channel.
For the $\chi_0\texttt{-}\chi_0$ case, we set $\sigma_0/m = 147.1~\rm cm^2/g$ and $w = 24.33~\rm km/s$, following the \texttt{VD100} model in Ref.~\cite{yang:2022mxl}, which yields a velocity-dependent cross section analogous to Ref.~\cite{turner:2020vlf}. See Supplemental Material for further details.

\begin{figure}[htbp]
  \centering
  \includegraphics[height=7.2cm]{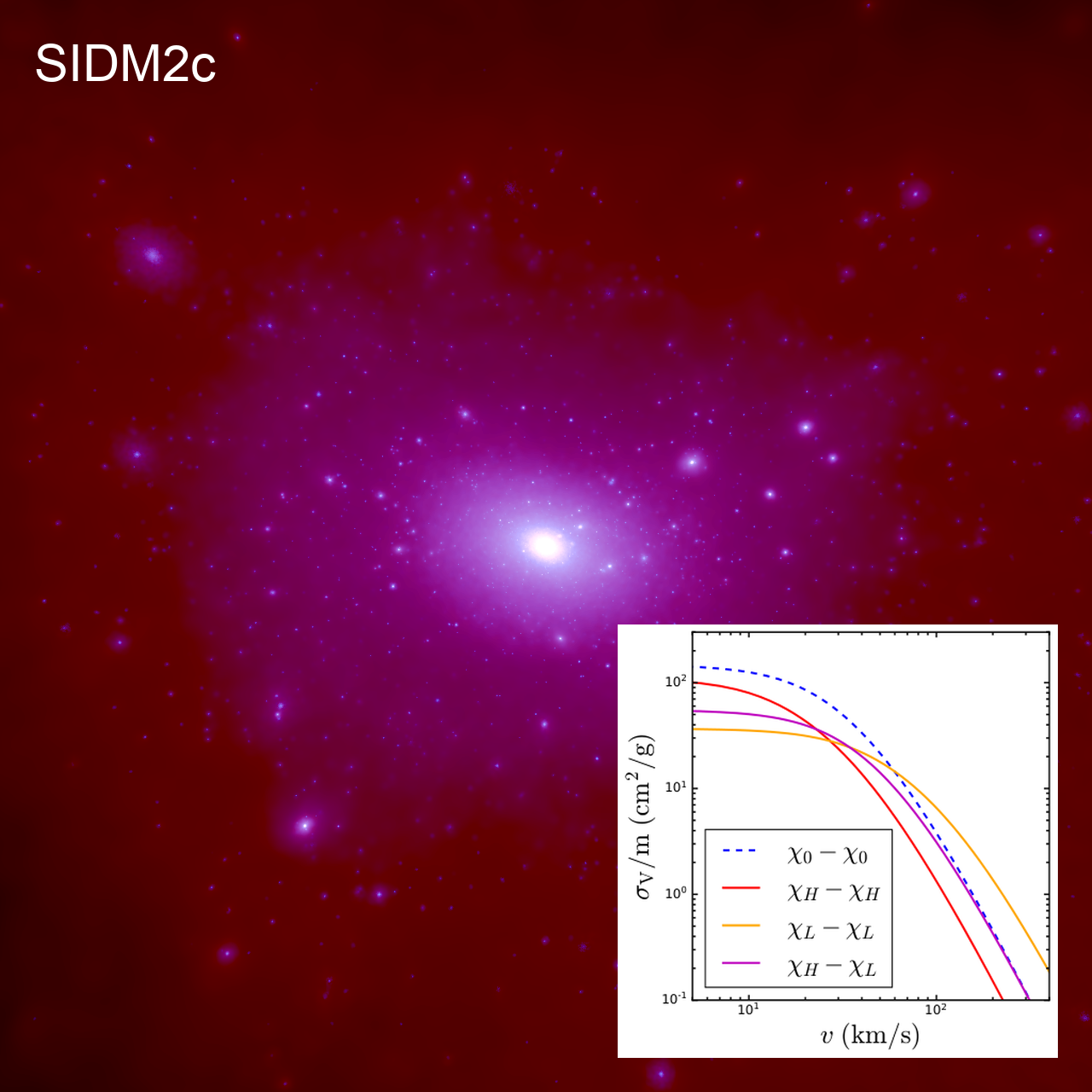}
  \caption{\label{fig:demo} 
Projected dark matter densities of MW analogs in the two-component \texttt{SIDM2c} simulation. 
Densities of the heavier (blue-to-white) and lighter (red-to-white) species are shown, with brighter regions indicating higher densities.
The inset shows viscosity cross sections~\cite{Tulin:2013teo} for the $\chi_H-\chi_H$ (red), $\chi_L-\chi_L$ (orange), and $\chi_H-\chi_L$ (magenta) interactions in the \texttt{SIDM2c} simulation. For comparison, the $\chi_0-\chi_0$ (blue) interaction in the one-component case is shown as a dashed curve. 
All channels assume the same mediator mass and coupling.
Model parameters for these interactions can be derived by fixing $\sigma_0/m = 147.1~\rm cm^2/g$ and $w = 24.33~\rm km/s$ in the $\chi_0\text{--}\chi_0$ case; see Supplemental Material for details. 
}
\end{figure}

We perform cosmological zoom-in simulations of the \texttt{AGORA} Milky Way (MW) analog system~\cite{AGORA:2013gut,AGORA:2021jss} using the \texttt{Gadget2} program~\cite{Springel0505010,Springel:2000yr}, adopting $\Omega_m = 0.272$, $\Omega_\Lambda = 0.728$, $h = 0.702$, and a box size of $L = 60~\rm Mpc/h$.  
The initial condition at $z = 100$ is generated using the \texttt{MUSIC} program~\cite{2011MNRAS.415.2101H}, with the transfer function computed by the \texttt{CAMB} program~\cite{2011ascl.soft02026L} based on these parameters. 
We randomly divide the initial particles into two components and conduct five cosmological simulations in both CDM and SIDM. These simulations include both single-component (\texttt{1c}) and two-component (\texttt{2c}) models, incorporating self- and cross-interaction (\texttt{SIDMx}) channels, as summarized in Table~\ref{tab1}.
The SIDM simulations are based on the module implemented in Refs.~\cite{yang220503392,yang:2022mxl}, which supports both types of differential cross sections considered in this work.
In the high-resolution region, the particle mass for the lighter component is $12 (1.5) \times 10^4~{\rm M_{\odot}}/h$ in SIDM (CDM) simulations, enabling us to probe structures down to a few times the softening length, $\epsilon = 0.16 (0.06)~{\rm kpc}/h$. 

\begin{table}[h]
    \centering
    \begin{tabular}{lccc}
        \hline
        Simulation & Components & Interaction Type & Color \\
        \hline
        \texttt{CDM1c}  & $\chi_0$      & CDM    & Green\\
        \texttt{SIDM1c} & $\chi_0$      & $\chi_0-\chi_0$ & Orange  \\
        \texttt{CDM2c}  & $\chi_H$, $\chi_L$  & CDM    & Blue \\
        \texttt{SIDM2c}   & $\chi_H$, $\chi_L$  & $\chi_{H,L}-\chi_{H,L}$ & Red \\
        \texttt{SIDMx}  & $\chi_H$, $\chi_L$  & $\chi_{H}-\chi_L$ & Magenta \\
        \hline
    \end{tabular}
    \caption{\label{tab1} Summary of the cosmological simulations performed in this work. The one-component (\texttt{1c}) simulations use a particle mass of $m_0 = (2/3)m_H$. The two-component \texttt{SIDM2c} simulations set $m_H = 3m_L$ and include both self- and cross-component interactions. The \texttt{SIDMx} simulation considers only cross-component scattering between $\chi_H$ and $\chi_L$.
}
\end{table}

Figure~\ref{fig:demo} presents the projected dark matter density of the MW analog in the \texttt{SIDM2c} simulation at $z=0$. The heavier and lighter species are shown in blue-to-white and red-to-white, respectively, with brighter (darker) regions indicating higher (lower) densities.  
The inset panel illustrates the velocity-dependent interactions in the two-component (solid) and one-component (dashed) models through the viscosity cross sections $\sigma_V/m$~\cite{Tulin:2013teo,Cline:2013pca,Boddy:2016bbu,Blennow:2016gde,Alvarez:2019nwt,Colquhoun:2020adl,yang220503392}. 
Compared to the one-component case, the two-component cross sections have lower normalization parameters $\sigma_0/m$ but larger velocity transition scales $w$. 
For quantitative analysis, we modify the \texttt{Rockstar} halo finder to accommodate two-component dark matter,  which enables identification and characterization of halos in simulation snapshots.
To facilitate density profile measurements, we restrict our analysis to MW subhalos with virial masses $M_{\rm vir} > 5\times 10^8~\rm M_{\odot}/h$, ensuring a minimum of approximately $2000$ particles per halo. 
Convergence at this resolution has been demonstrated in Appendix B of Ref.~\cite{yang:2022mxl}. We adopt $214\ \rm kpc/h$ as an estimate of the virial radius for MW analogs across all simulations.

\begin{figure*}[htbp]
  \centering
  \includegraphics[width=5.5cm]{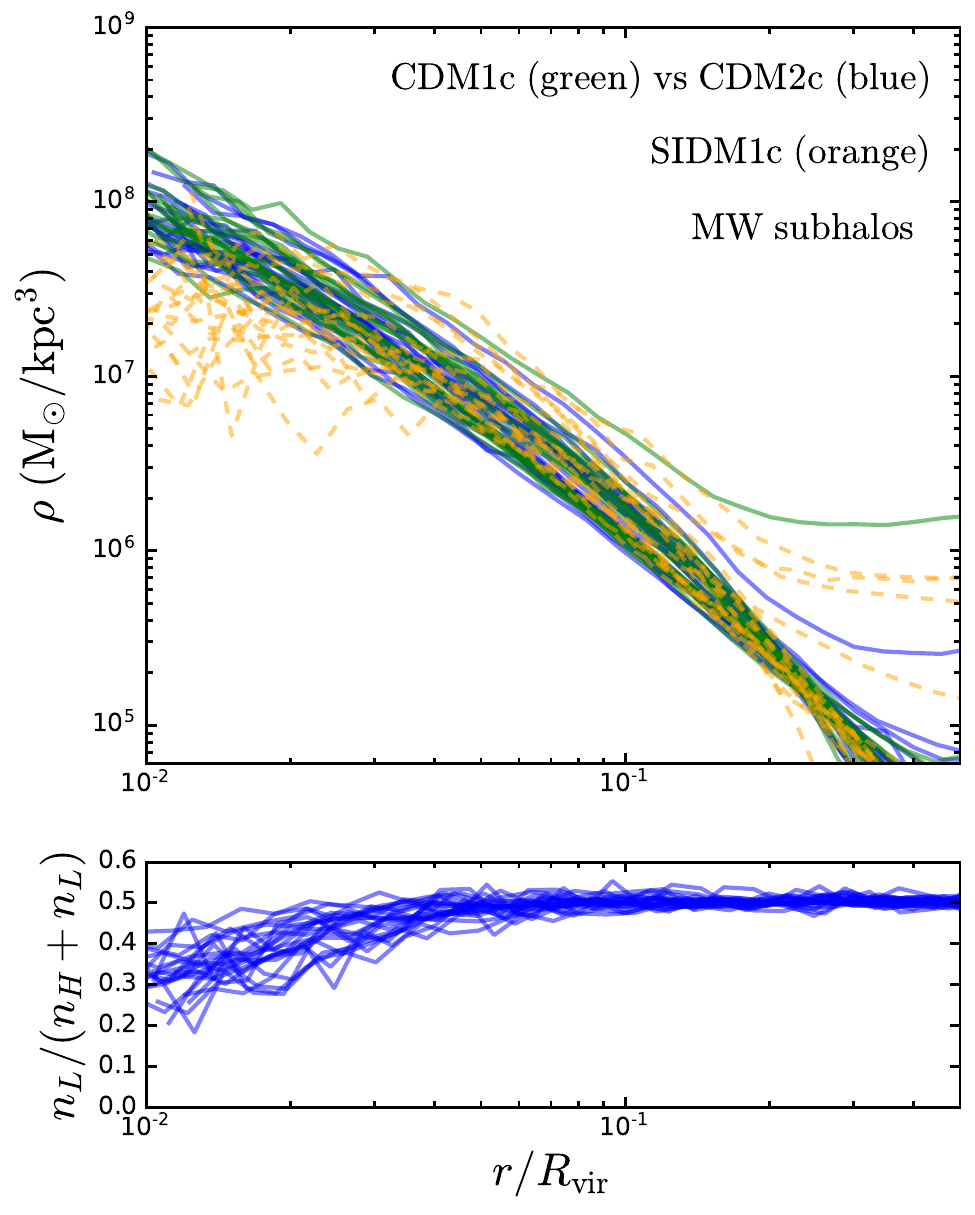}
  \includegraphics[width=5.5cm]{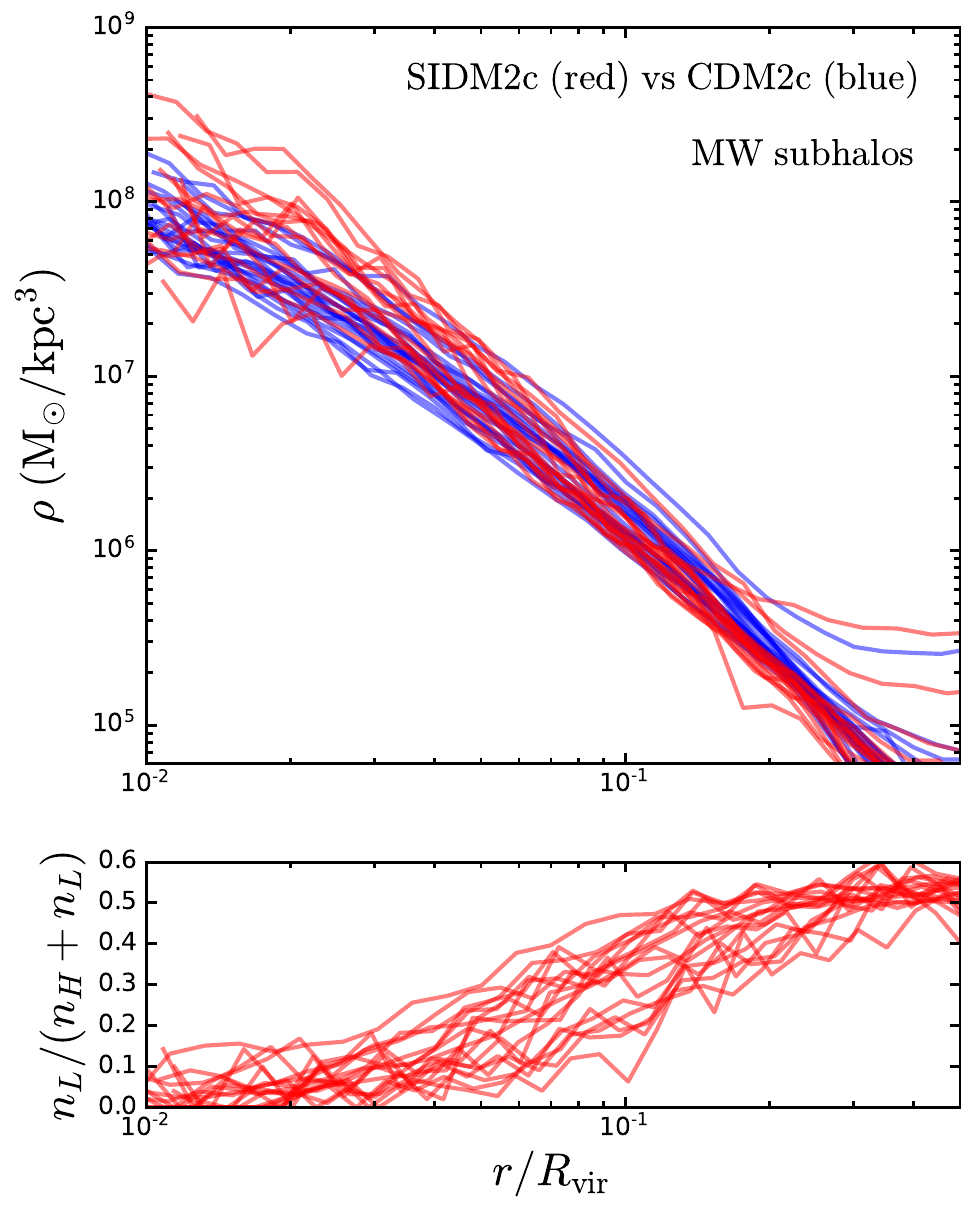}
  \includegraphics[width=5.5cm]{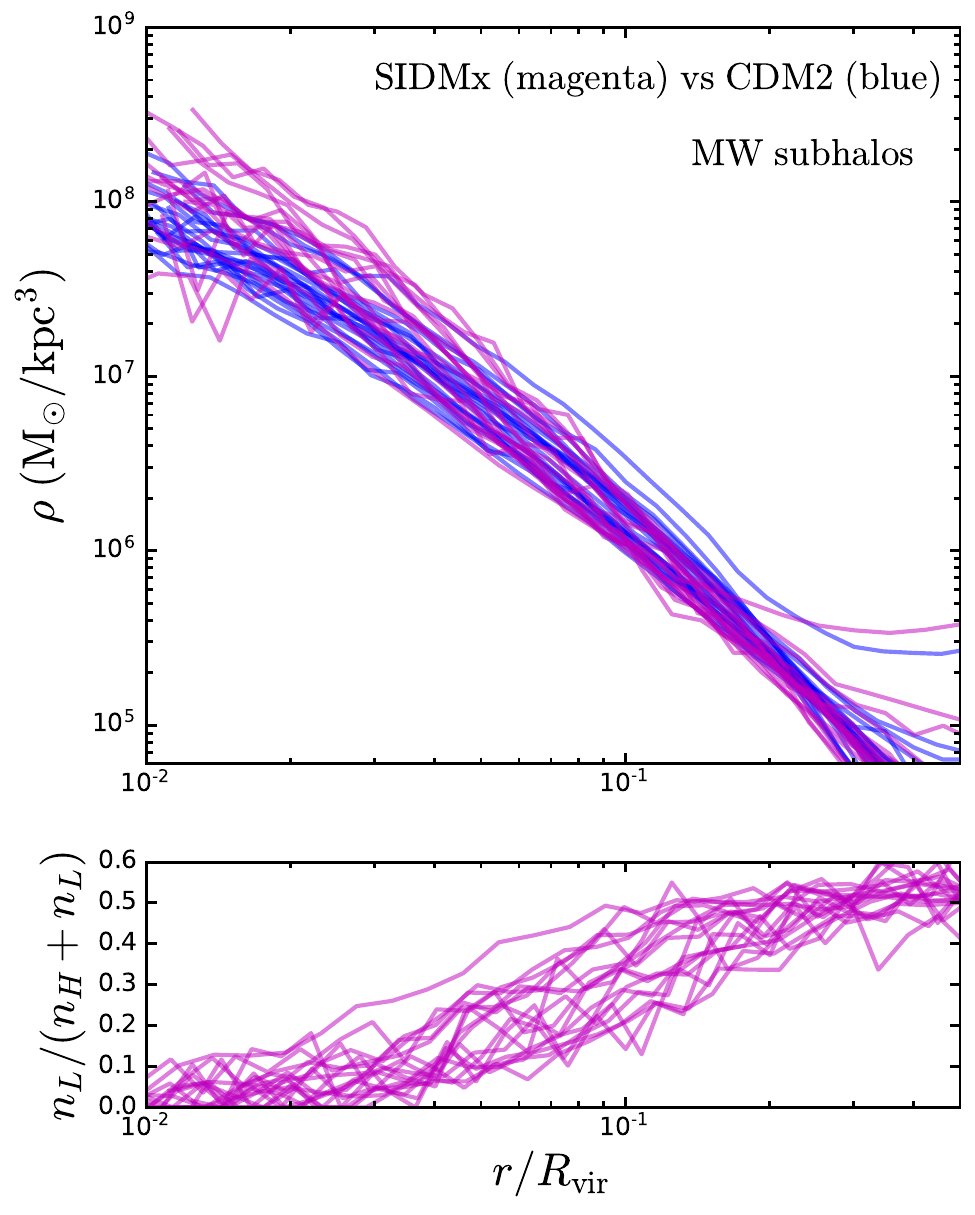}
  \caption{\label{fig:profs} 
Halo density profiles of MW subhalos of $M_{\rm vir}>5\times 10^8~\rm M_{\odot}/h$ in simulated models. 
The main panels show \texttt{CDM1c} (green) and \texttt{SIDM1c} (orange) on the left panel, \texttt{SIDM2c} (red) and \texttt{SIDMx} (magenta) on the middle and right panels, respectively. The \texttt{CDM2c} (blue) is shown in all panels for comparison. 
The smaller sub-panels display the fractional number density of the lighter species, $n_L/(n_H+n_L)$. While the \texttt{CDM2c} case closely resembles the \texttt{CDM1c} scenario, with only slightly shallower inner regions, SIDM significantly redistributes the lighter species to larger radii, leading to denser inner regions dominated by the heavier species. In the \texttt{SIDM2c} case, self-interactions among the heavier species introduce both core and cusp profiles.
}
\end{figure*}

{\it Mass segregation and diversity.}
In the two-component scenario, interactions between species of different masses tend to equalize their kinetic energies, causing the more massive species to sink into the inner halo regions---an effect known as mass segregation.  
In CDM, this effect is governed by the relaxation time due to gravitational scatterings, which scales as approximately $N /(10\ln N)$ times the particle crossing time. Since the particle number $N$ is extremely large, the relaxation time can be exceedingly long. 
In contrast, mass segregation by SIDM proceeds via collisional relaxation, with a timescale independent of $N$ and given by $1 / (\rho \sigma v)$, where $\rho$ is the local density, $v$ the relative velocity, and $\sigma$ the scattering cross section. A larger $\sigma$ thus accelerates relaxation and enhances mass segregation.

Based on the particle positions and masses from the modified \texttt{Rockstar} program~\cite{Behroozi:2011ju}, we compute the density profiles of subhalos in MW analogs and show them in Fig.~\ref{fig:profs} with radii normalized by the halos' virial radii.
The subpanels display the radial distribution of the lighter species, quantified by $n_L/(n_H+n_L)$, as a measure of mass segregation.
Compared to \texttt{CDM2c} (left), both \texttt{SIDM2c} (middle) and \texttt{SIDMx} (right) show a clear expulsion of the lighter species to larger radii, with its central fraction dropping to nearly zero.
The nearly identical segregation patterns in \texttt{SIDM2c} and \texttt{SIDMx} confirm that cross-component interactions primarily drive this effect.
Meanwhile, the agreement between \texttt{CDM1c} and \texttt{CDM2c} density profiles demonstrates the negligible impact of mass segregation in CDM.

To facilitate comparison with the one-component case, we include \texttt{SIDM1c} density profiles as dashed curves in the left panel. 
For the chosen cross section, 
\texttt{SIDM1c} develops prominent cores with central densities falling below observational estimates (e.g., Ref.~\cite{kaplinghat:2019svz}). In contrast, mass segregation in the two-component models promotes the formation of denser central regions in many subhalos, resembling the core-collapse phase seen in single-component SIDM. As a result, both \texttt{SIDM2c} and \texttt{SIDMx} retain cuspy profiles.
In \texttt{SIDM2c}, self-interactions among heavy particles further enhance structural diversity, producing both cores and cusps. As we will show, this variation in inner densities enables \texttt{SIDM2c} to better reproduce observational data.

The interplay between mass segregation and self-interactions thus broadens the phenomenology of SIDM. In particular, segregation allows some halos to exhibit core-collapsing features even at modest cross sections where one-component models would form overly large cores. 
This behavior can be traced to the maintenance of pressure equilibrium during cross-component scatterings. 
A decrease in the velocity of the heavier component, $v_H$, must be compensated by an increase in its density to keep the pressure, $\propto \rho_H v_H^2$, constant.The extent of this effect depends on multiple factors: the mass difference and cross-component scattering rates determine the degree of mass segregation, while the self-interaction cross section of the heavier component regulates the diversity in density profiles. 
These effects intertwine, collectively reshaping small-scale structures in the inner halo regions. 
In the limit where the two particle species have equal masses, mass segregation disappears, and the two-component model effectively reduces to the one-component SIDM scenario, although cross-species interactions remain governed by a Rutherford-like scattering cross section.

\begin{figure}[htbp]
  \centering
  \includegraphics[height=7.cm]{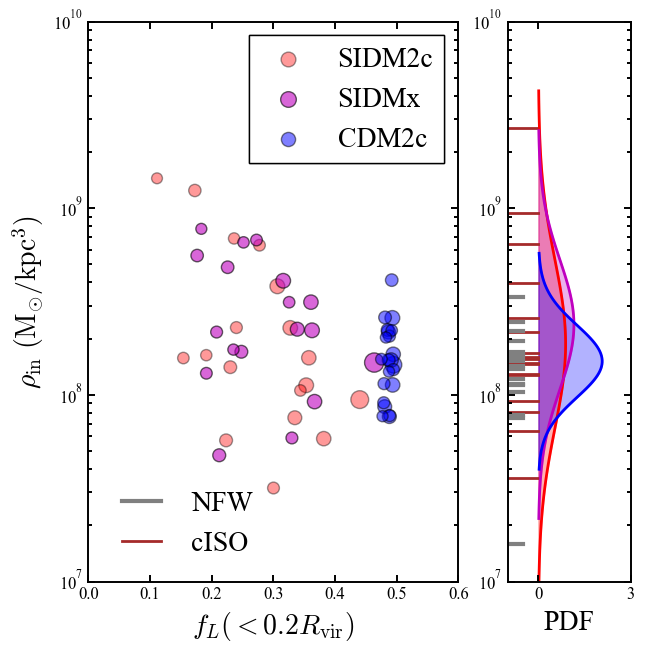}
  \caption{\label{fig:rhoinn2}
Extrapolated inner halo density, $\rho_{\rm in}=\rho(r=150{\ \rm pc})$, vs fractional number density, $f_L(<0.2 R_{\rm vir})$, for MW subhalos of $M_{\rm vir}>5\times 10^8~\rm M_{\odot}/h$ in two-component simulations \texttt{SIDM2c} (red), \texttt{SIDMx} (magenta), and \texttt{CDM2c} (blue). 
For comparison, the $\rho_{\rm in}$ of observed MW satellites, derived from Ref.~\cite{kaplinghat:2019svz} under the assumptions of NFW (gray) and isothermal (brown) profiles, are shown as tick marks along the y-axis, with their Gaussian fits overlaid to highlight the probability distributions.
The size of the points is proportional to the virial radius of the corresponding halos.
}
\end{figure}

{\it Inner halo densities and the fraction of lighter dark matter.}
In SIDM, gravothermal evolution has been identified as a key mechanism driving diversity in inner halo structures. Here, we demonstrate that cross-component scattering introduces an additional source of diversity via mass segregation, which can act in conjunction with the intra-component gravothermal evolution. To quantify this effect, we evaluate the inner halo density, $\rho_{\rm in} \equiv \rho(r = 150~\rm pc)$, and the fractional number density of the lighter component, presenting the distributions for MW subhalos in Fig.~\ref{fig:rhoinn2}. 
As the radius of $150$~pc approaches the softening length in our simulations, we fit the density profiles using a parametric model to obtain more robust results~\cite{yang:2023jwn}; see the Supplemental Material for details. 
In the literature, $\rho_{\rm in}$ has been reconstructed under various density profile models to test the diversity of observed MW satellites~\cite{Andrade:2023fgr,kaplinghat:2019svz,Read:2018fxs,Read:2018pft,Hayashi:2020jze}. The fractional number density of the lighter component, defined as
$f_L(<r) \equiv n_L(<r)/[n_H(<r) + n_L(<r)]$,
and evaluated within $0.2 R_{\rm vir}$, enables us to further probe the correlation between mass segregation and halo diversity. This provides a means to quantify cross-component SIDM as a new source of structural diversity.

To assess how the diversity in our simulations compares with observations, we use data compiled in Ref.~\cite{kaplinghat:2019svz}, which analyzes MW satellite galaxies, including both classical dwarfs and ultrafaints, and extracts inner halo densities at $150$~pc using two extrapolation methods: one assuming Navarro-Frenk-White (NFW) profiles~\cite{1997ApJ...490..493N} and the other assuming cored isothermal (cISO) profiles~\cite{Kaplinghat:2015aga}, the latter intended to model SIDM halos. 
We represent these extrapolated densities as horizontal ticks along the $y$-axis, with NFW-based values in gray and cISO-based values in brown. 
To compare data with our simulation results, we project the $\rho_{\rm in}$ of the selected subhalos onto the right panel, fitting their values in log space and display the resulting probability distributions as shaded Gaussians along the $y$-axis, using the same color scheme as the simulation benchmarks. 

In the main panel of Fig.~\ref{fig:rhoinn2}, we show the distribution of $\rho_{\rm in}$ vs $f_L(<0.2\,R_{\rm vir})$ for the three two-component simulations. In the \texttt{CDM2c} (blue) case, all subhalos have $f_L$ values clustered near $0.5$, whereas in \texttt{SIDM2c} and \texttt{SIDMx}, $f_L$ spans nearly the full range from $0$ to $0.5$, reflecting substantial variation in mass segregation across halos.

The \texttt{SIDMx} case includes only cross-component scatterings, and thus lacks gravothermal evolution in the conventional sense. 
Nonetheless, we observe a broader $\rho_{\rm in}$ distribution with a median shifted upward relative to \texttt{CDM2c}, indicating that cross-component interactions alone can generate significant diversity in inner halo structure.

When self-interactions are introduced among the same components (\texttt{SIDM2c}), the scatter further increases, yielding a larger population of core-like halos and a slightly lower median $\rho_{\rm in}$. 
Interestingly, the resulting diversity is consistent with \texttt{CDM2c} when inner densities are extrapolated using NFW profiles, but aligns more closely with the two SIDM cases under the assumption of cISO profiles. This highlights that current observational uncertainties limit our ability to discriminate between CDM and SIDM scenarios. Improved measurements and larger satellite samples will be essential to resolve the inner structure diversity in MW satellites.

Apart from the overall scatter, we find a moderate anti-correlation between $\rho_{\rm in}$ and $f_L(<0.2\,R_{\rm vir})$. In \texttt{SIDMx}, the Pearson correlation coefficient is $r = -0.36$ with a p-value of $0.14$, indicating a weak but coherent trend. The anti-correlation becomes stronger in \texttt{SIDM2c}, with $r = -0.60$ and $p = 0.008$, suggesting that increased mass segregation is associated with enhanced central densities.

{\it Discussion and conclusion.}
We have investigated the impact of mass segregation in two-component SIDM models and its implications for the diversity of inner halo structures.
Mass segregation, a well-known phenomenon in astrophysical systems involving globular clusters and black holes, can also arise in multicomponent dark matter scenarios.
In CDM, the segregation timescale is too long to yield observable consequences. By contrast, cross-component scatterings in SIDM can significantly accelerate this process, leading to substantial reshaping of halo density profiles.

Our results show that SIDM naturally connects mass segregation to structural diversity. In particular, we find that segregation can drive a subset of halos toward higher central densities, resembling core-collapse-like configurations. This behavior complements standard gravothermal evolution and introduces an additional mechanism for generating diversity in inner halo densities.
While diversity in single-component SIDM is typically interpreted as a signature of gravothermal evolution, our findings highlight its potential as a probe of dark sector composition. Precise characterization of inner halo structures may offer clues to the presence of multiple self-interacting components, opening a new observational window into the particle nature of SIDM.

Although our analysis focuses on MW–like halos, the underlying mechanism is general and may operate in more massive systems. In such cases, dense, segregated substructures may leave observable imprints, such as enhanced gravitational lensing. More broadly, our results suggest a novel approach for resolving potential tensions in SIDM by enabling shifts in central densities without reducing the overall diversity.

Note added: After the submission of this work, Ref.~\cite{Patil:2025nmj} explored halo evolution under cross-component interaction with mass segregation.
Ref.~\cite{Yang:2025xsp} further elaborated on observational signatures from SIDM with mass segregation in dwarf halos and cluster substructures.
The findings of these works align and complement. 

\acknowledgments
We thank Hai-Bo Yu, Haipeng An, and Xiaoyong Chu for helpful discussion. 
The authors were supported in part by the National Key Research and Development Program of China (No. 2022YFF0503304), the Project for Young Scientists in Basic Research of the Chinese Academy of Sciences (No. YSBR-092).

\textbf{Software:} \texttt{Rockstar}~\cite{Behroozi:2011ju}, \texttt{Scipy}~\cite{2020SciPy-NMeth}, \texttt{NumPy}~\cite{5725236}, \texttt{Matplotlib}~\cite{4160265}.

\bibliography{References}

\appendix

\section{Implementation of differential scatterings}

We consider a simple dark QED-like model with two dark matter species, $\chi_H$, $\chi_L$ of mass $m_H$ and $m_L$, interacting via a massive vector mediator of mass $m_V$ and coupling $e_D$~\cite{Kaplan:2009ag,Petraki:2011mv,Petraki:2013wwa,Petraki:2014uza,ZUREK201491}. The corresponding dark fine-structure constant is defined as $\alpha_D = e_D^2 / (4\pi)$. We work in a weakly coupled perturbative regime, where differential cross sections can be calculated at tree level. To describe the interactions among $\chi_L$ or $\chi_H$ (self-interactions) and between $\chi_L$ and $\chi_H$ (cross-component interactions), we adopt the parameterizations of the Møller and Rutherford cross sections from Ref.~\cite{yang220503392}.

Rutherford scattering applies to the interaction of two distinct species. When the masses of the two particles differ, the center-of-mass frame differential cross section can be generalized as~\cite{Feng:2009hw,Ibe:2009mk}
\begin{eqnarray}
\label{eq:xsr}
\frac{d\sigma}{d \cos\theta} = \frac{\sigma_{0}w^4}{2\left[w^2+{v^{2}}\sin^2(\theta/2)\right]^2 },
\end{eqnarray}
where $\sigma_0\equiv \pi\alpha^2_D/(\mu^2 w^4)$, $w= m_V/(2\mu)$, and $\mu=m_L m_H/(m_L+m_H)$ is the reduced mass of the two-particle system. Here, $v$ is the relative velocity and $\theta$ is the scattering angle. Integrating over $\theta \in (0,\pi)$ gives the total cross section
\begin{equation}
\sigma_{\rm tot}= {\sigma_0}/(1+{v^2}/{w^2}),
\end{equation}
which we use to determine scattering probabilities in our simulations. For further details, see Ref.~\cite{yang220503392}.

Møller scattering describes the scattering of identical particles ($\chi\chi \rightarrow \chi\chi$). The differential cross section is~\cite{yang220503392,Girmohanta:2022dog}
\begin{equation}
\label{eq:xsm}
\dfrac{d\sigma}{d \cos\theta} = \frac{2\sigma_0 w^4 \left[\left(3 \cos^2\theta+1\right) v^4+4 v^2 w^2+4  w^4\right]}{\left(\sin^2\theta v^4+4 v^2 w^2+4 w^4\right)^2}, 
\end{equation}
where $\theta$ is restricted to $(0,\pi/2)$. The factor of $2$ ensures the total cross section matches that in Ref.~\cite{yang220503392}. Upon integration, one obtains
\begin{equation}
\sigma_{\rm tot}= \sigma_0 w^4 \left[\frac{1}{v^2 w^2+w^4}+\frac{1}{v^4+2   v^2 w^2} \ln \left(\frac{w^2}{v^2+w^2}\right)\right].
\end{equation}
In our simulations, we include an additional $1/2$ phase space factor to account for the identical nature of the two initial-state particles.

In this work, we fix the mediator mass $m_V$, the coupling constant $\alpha_D$, and the total particle number across different simulation types, which result in varying $\sigma_0/m$ and $w$ parameters. 
In the single-component case, we denote the dark matter particles by $\chi_0$ and set their mass to $m_0 = (2/3)m_H$.
Table~\ref{tab2} summarizes the corresponding SIDM model parameters for each type of scattering. 
\begin{table}[h]
    \centering
    \begin{tabular}{c|ccc}
        \hline
        Scatterings & Mass relation & $\frac{\sigma_0}{m}$ relation & $w\ \rm(km/s)$ relation  \\
        \hline
        $\chi_0{\texttt-} \chi_0$  & $m_0=\frac{2}{3} m_H$ & $\frac{\sigma_M}{m_0}=147.1\ \rm \frac{cm^2}{g}$ & $w_M=24.33\ \rm \frac{km}{s}$  \\
        \hline
        $\chi_H-\chi_H$  & $m_H=\frac{3}{2} m_0$ & $\frac{\sigma_H}{m_H}=\frac{3}{2}\frac{\sigma_M}{m_0}$ & $w_1 = \frac{2}{3} w_M$ \\ 
        $\chi_L-\chi_L$  & $m_L= \frac{1}{3}m_H$ & $\frac{\sigma_L}{m_L}=\frac{1}{3}\frac{\sigma_H}{m_H}$ & $w_2 = 3 w_1$ \\ 
        $\chi_H-\chi_H$  & $m_H=3 m_L$ & $\frac{\sigma_x}{m_H}=\frac{1}{4}\frac{\sigma_H}{m_H}$ & $w_x = 2 w_1$ \\
        \hline
    \end{tabular}
    \caption{\label{tab2} SIDM model parameters for different scattering processes in a dark QED theory with fixed $m_V$ and $\alpha_D$. The mass assignments ensure the total particle number remains the same in all simulations.
}
\end{table}

\begin{figure*}[htbp]
  \centering
  \includegraphics[width=5.5cm]{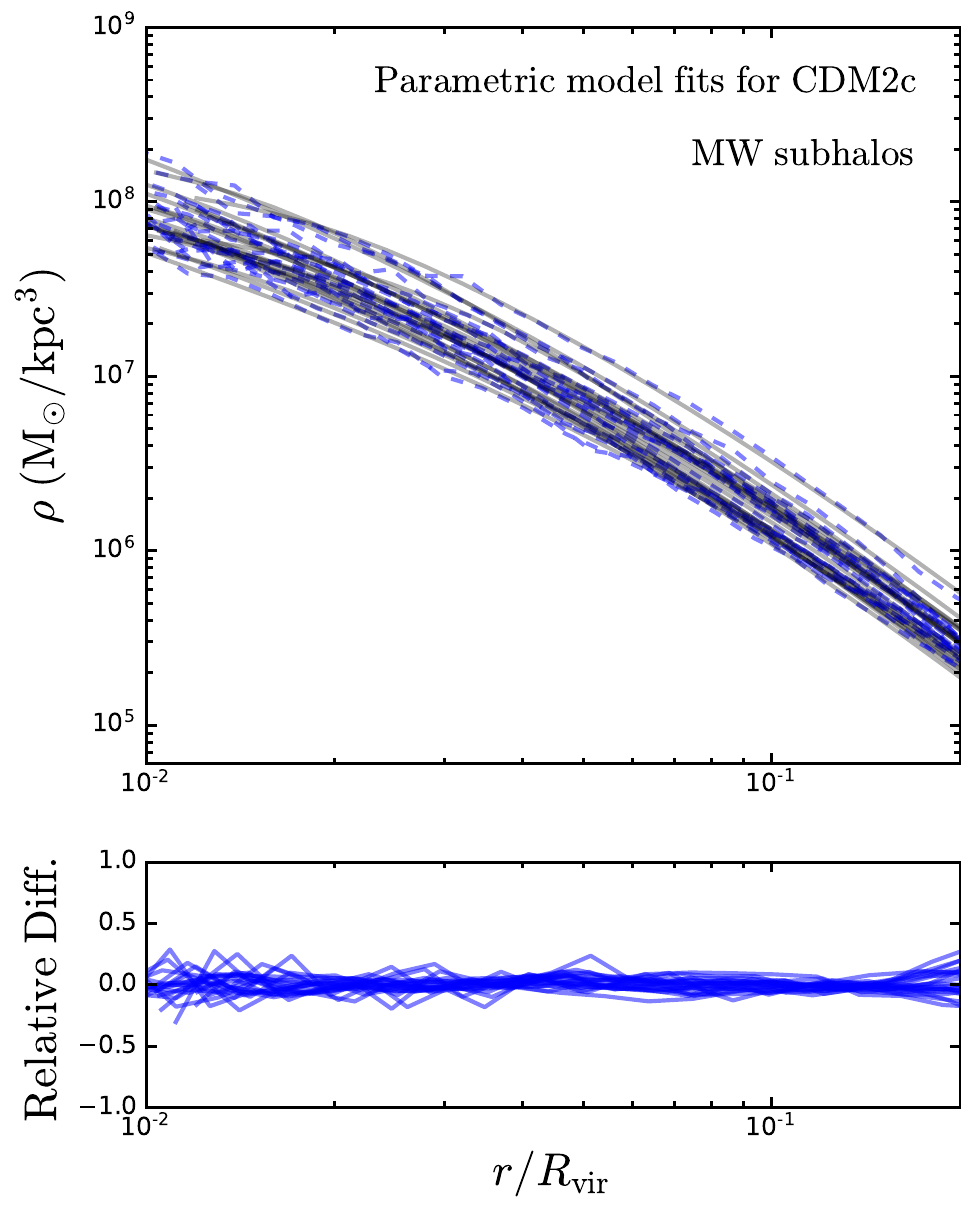}
  \includegraphics[width=5.5cm]{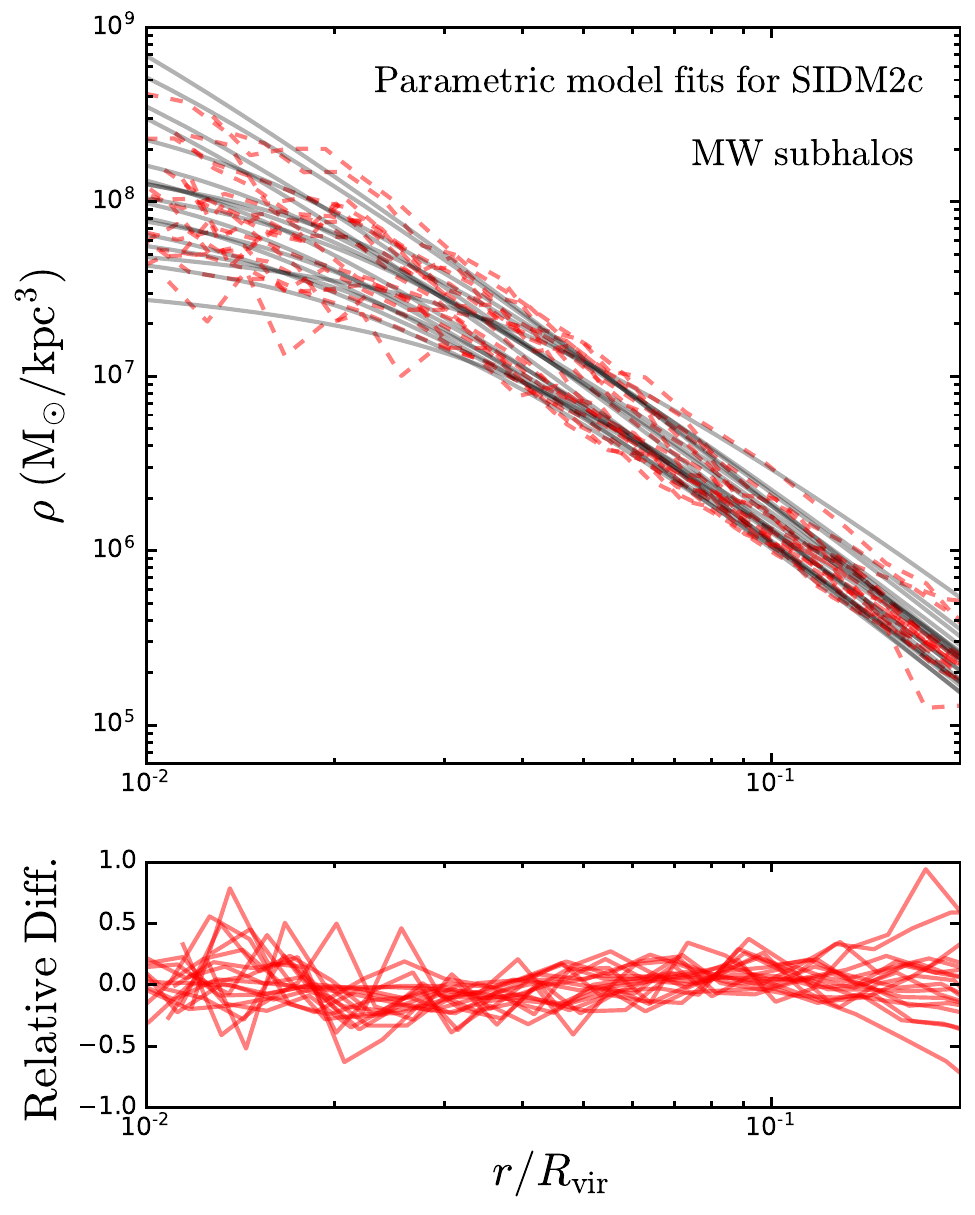}
  \includegraphics[width=5.5cm]{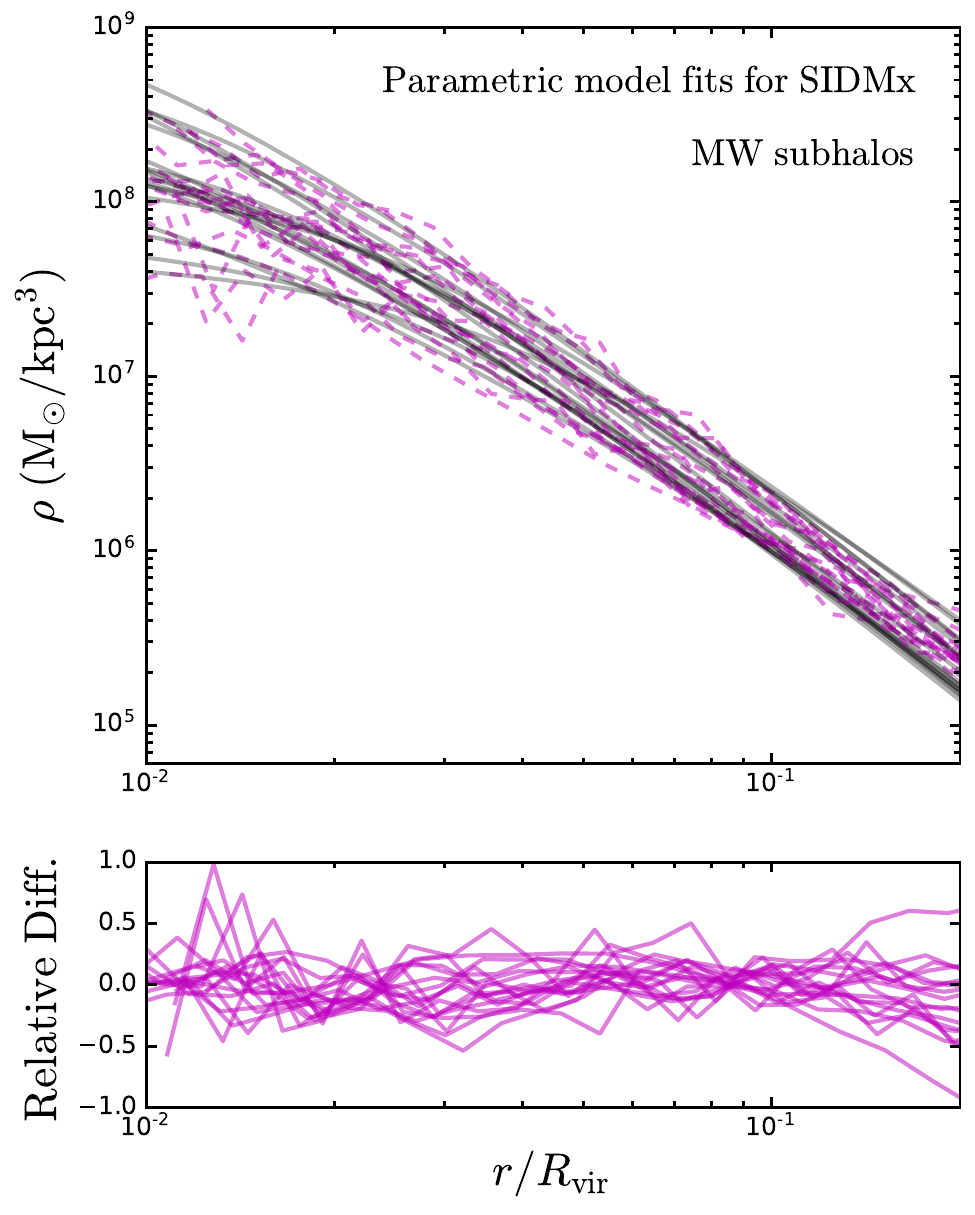}
  \caption{\label{fig:fits}
The parametric model fitting results (solid black) for the density profiles of MW subhalos with $M_{\rm vir}>5 \times 10^8~\rm M_{\odot}/h$ in \texttt{CDM2c} (left), \texttt{SIDM2c} (middle), and \texttt{SIDMx} (right) simulations. 
The relative differences between the simulated (Sim) and fitted (Fit) curves are measured as $2\rm (Fit-Sim)/(Fit+Sim)$ and shown in the sub-panels. Compared with the profiles in the main article, more bins are used here to improve the fit. The fitted curves are used to obtain the $\rho_{\rm in}=\rho(r=150{\ \rm pc})$, as referenced in the main text. 
}
\end{figure*}

The viscosity cross section $\sigma_V$ is calculated by weighting the differential cross section with a kernel appropriate for kinetic-theory studies of viscosity and heat conductivity~\cite{Tulin:2013teo}:
\begin{equation}
\label{eq:xsrD}
\sigma_V = \frac{3}{2} \int d\cos\theta \sin^2\theta \frac{d\sigma}{d\cos\theta}.
\end{equation}
This quantity is of particular interest for studying the transport properties of SIDM halos and is used to generate the inset of Fig.~1 in the main text.
See Ref.~\cite{yang220503392} for analytic results for the Møller and Rutherford scattering cases. 

\section{Fitting the density profiles}

The radius of $150$ pc, at which we evaluate the inner halo densities $\rho_{\rm in}$, lies below the softening length of the simulations. To obtain more reliable estimates of $\rho_{\rm in}$, we fit the simulated density profiles and extract values from the resulting fits.

We find that the parametric profile commonly used to model one-component SIDM halos~\cite{yang:2023jwn,Yang:2024tba,Yang:2024uqb,Hou:2025gmv} can still be used to describe halos in two-component dark matter. Specifically, we fit the density profiles using the following $\beta4$ profile: 
\begin{eqnarray}
\label{eq:cnfw}
\rho_{\rm \beta4}(r) = \frac{\rho_s}{\frac{\left(r^{4}+r_c^{4} \right)^{1/{4}}}{r_s} \left(1 + \frac{r}{r_s} \right)^2}, 
\end{eqnarray}
where $\rho_s$, $r_s$, and $r_c$ are free parameters that evolve with the normalized time variable $\tau \equiv t/t_c$. We treat $\tau$, $V_{\rm max}$, and $R_{\rm max}$ as fitting parameters and convert them to $\rho_s$, $r_s$, and $r_c$ using the relations provided by the parametric model for SIDM halos in Ref.~\cite{yang:2023jwn}.
Specifically, given $\tau$, $V_{\rm max}$, and $R_{\rm max}$, we first evaluate 
\begin{eqnarray}
\hat{V}_{\rm max} &=& 1+ 0.1777 \tau -4.399 \tau^3 + 16.66 \tau^4 - 18.87 \tau^5 \\ \nonumber 
 && + 9.077 \tau^7 - 2.436 \tau^9  \\ 
\hat{R}_{\rm max} &=& 1 + 0.007623 \tau - 0.7200 \tau^2 + 0.3376 \tau^3 \\ \nonumber
 && -0.1375 \tau^4,
\end{eqnarray}
obtaining NFW scale parameters as 
\begin{eqnarray}
r_{s,0} &=& \frac{R_{\rm max}}{\hat{R}_{\rm max} 2.1626}, \\ \nonumber
\rho_{s,0} &=&\frac{1}{G} \left( \frac{V_{\rm max}}{1.648 r_{s,0}\hat{V}_{\rm max}} \right)^2. 
\end{eqnarray} 
Based on the $\rho_{s,0}$, $r_{s,0}$, and $\tau$, the three parameters in the $\beta4$ profile is evaluated as
\begin{eqnarray}
\frac{\rho_s}{\rho_{s,0}} &=& 2.033 + 0.7381 \tau + 7.264 \tau^5  \\ \nonumber 
&& -12.73 \tau^7 + 9.915 \tau^9  \\ \nonumber
&&+ (1-2.033) (\ln 0.001)^{-1} \ln \left( \tau + 0.001 \right), \\ \nonumber 
\frac{r_s}{r_{s,0}} &=& 0.7178 - 0.1026 \tau +  0.2474 \tau^2 -0.4079 \tau^3 \\ \nonumber
&& + (1-0.7178) (\ln 0.001)^{-1} \ln \left( \tau + 0.001 \right), \\ \nonumber 
\frac{r_c}{r_{s,0}} &=& 2.555 \sqrt{\tau} -3.632 \tau + 2.131 \tau^2 -1.415 \tau^3 \\ \nonumber
&& + 0.4683 \tau^4.
\end{eqnarray}

Figure~\ref{fig:fits} presents the fitted (Fit) and simulated (Sim) density profiles for the \texttt{CDM2c}, \texttt{SIDM2c}, and \texttt{SIDMx} simulations, from left to right. The relative differences, measured as $2{\rm (Fit - Sim)/(Fit + Sim)}$, are shown in the subpanels and indicate good agreement across all cases. While the relative differences occasionally reach $\sim 50\%$, these fluctuations are driven by statistical noise in the SIDM simulations and do not exhibit any systematic bias.
\end{document}